\documentclass[twocolumn]{openjournal}

\usepackage{lipsum}

\usepackage{xcolor}
\usepackage{textgreek}
\usepackage[utf8]{inputenc}
\usepackage[english]{babel}
\usepackage{xspace}
\usepackage{amsmath}

\usepackage{hyperref}
\hypersetup{
    unicode, 
    colorlinks=true,
    linkcolor=linkcolor,
    citecolor=linkcolor,
    filecolor=linkcolor,
    urlcolor=linkcolor,
}
\usepackage{color,colortbl}
\definecolor{linkcolor}{rgb}{0.0,0.3,0.5}
\usepackage{tensind}
\tensordelimiter{?}
\DeclareGraphicsExtensions{.bmp,.png,.jpg,.pdf}
\usepackage{verbatim}
\usepackage[normalem]{ulem}
\usepackage{orcidlink}
\usepackage{soul}

\graphicspath{ {./figs/} }

\begin{document}
\title{Revealing the companion of NGC 1850-BH1}

\def\bh1{NGC~1850-BH1}
\newcommand{\Teff}{\ensuremath{T_\mathrm{eff}}}
\newcommand{\Logg}{\ensuremath{\log{g}}}
\providecommand{\msun}{\ensuremath{\mathrm{M}_\odot}}
\providecommand{\rsun}{\ensuremath{\mathrm{R}_\odot}}
\newcommand{\vsini}{\ensuremath{v\sin(i)}}
\newcommand{\chisq}{\ensuremath{\chi^2}}

\newcommand{\TeffA}{10650}
\newcommand{\TeffB}{18660}
\newcommand{\loggA}{3.0}
\newcommand{\loggB}{3.8}
\newcommand{\vsiniA}{68}
\newcommand{\vsiniB}{275}
\newcommand{\radA}{6.7}
\newcommand{\radB}{4.1}
\newcommand{\massA}{1.7}
\newcommand{\massB}{5.2}
\newcommand{\disq}{3.3}
\newcommand{\spectroq}{2.6}

\providecommand{\mass}[1]{\ensuremath{M_{\mathrm{#1}}}}
\providecommand{\logg}[1]{\ensuremath{\log{g}_{\mathrm{#1}}}}
\providecommand{\radius}[1]{\ensuremath{R_{\mathrm{#1}}}}
\providecommand{\teff}[1]{\ensuremath{T_{\mathrm{eff, #1}}}}
\providecommand{\vel}[1]{\ensuremath{v_{\mathrm{#1}}}}

\author{Rhys Seeburger\orcidlink{0000-0000-0000-0000}}
\email{l.m.seeburger@ljmu.ac.uk}
\affiliation{Astrophysics Research Institute, Liverpool John Moores University, IC2 Liverpool Science Park, 146 Brownlow Hill, Liverpool L3~5RF, UK}

\author{Sara Saracino\orcidlink{0000-0000-0000-0000}}
\email{sara.saracino@inaf.it}
\affiliation{INAF - Osservatorio Astrofisico di Arcetri, Largo E. Fermi 5, 50125 Firenze, Italy}
\affiliation{Astrophysics Research Institute, Liverpool John Moores University, IC2 Liverpool Science Park, 146 Brownlow Hill, Liverpool L3~5RF, UK}

\author{Sebastian Kamann\orcidlink{0000-0001-6604-0505}}
\email{s.kamann@ljmu.ac.uk}
\affiliation{Astrophysics Research Institute, Liverpool John Moores University, IC2 Liverpool Science Park, 146 Brownlow Hill, Liverpool L3~5RF, UK}

\author{Nate Bastian\orcidlink{0000-0001-5679-4215}}
\email{nate.bastian@dipc.org}
\affiliation{Donostia International Physics Center (DIPC), Paseo Manuel de Lardizabal, 4, 20018, Donostia-San Sebastián, Guipuzkoa, Spain; IKERBASQUE, Basque Foundation for Science, 48013, Bilbao, Spain}

\author{Tomer Shenar \orcidlink{0000-0003-0642-8107}}
\email{tshenar@tau.ac.il}
\affiliation{
The School of Physics and Astronomy, Tel Aviv University, Tel Aviv 6997801, Israel}

\author{Debasish Dutta \orcidlink{0009-0000-1383-8011}}
\email{debasish.dutta@ista.ac.at}
\affiliation{
Institute of Science and Technology Austria (ISTA),  Am Campus 1, 3400 Klosterneuburg, Austria}

\author{Julia Bodensteiner \orcidlink{0000-0002-9552-7010}}
\email{j.bodensteiner@uva.nl}
\affiliation{Anton Pannekoek Institute Institute for Astronomy, University of Amsterdam, Science Park 904, 1098 XH Amsterdam, The Netherlands}

\begin{abstract}
The nature of the system \bh1 is discussed controversially in the literature. Located in the young (100~Myr) massive cluster NGC~1850 in the Large Magellanic Cloud, it was initially proposed to consist of a 5 \msun\ main sequence turn-off star orbiting a stellar-mass black hole. However, subsequent analyses of the system have presented an alternative explanation, comprising of a lower mass (1-2 \msun) puffed-up donor star orbiting a rapidly rotating mass accretor. Efforts to discern the secondary's nature as a luminous star or a compact object in the original spectroscopy have remained inconclusive.
In this work, we analyse a set of follow-up multi-epoch FLAMES-ARGUS spectra, which are bluer and of higher spectral resolution than the original data. Exploiting the spectral disentangling technique, we reveal the luminous secondary, and constrain the parameters of both components. We can describe the system self-consistently, adopting a 1.7 $\pm$ 0.4 \msun\ puffed-up mass donor and a rapidly rotating (\vsini = \vsiniB~km/s) 5.2 $\pm$ 2.6 \msun\ luminous companion. While this rules out the black hole hypothesis, tensions with the analysis of the original data remain: Our inferred flux ratio of $\sim$ 35-50\% in the red optical wavelength range is higher than the 10\% upper limit found in previous work. The origin of this discrepancy remains unclear. Thanks to its location in a cluster of well-known mass and metallicity and its peculiar orbital properties, \bh1 has the potential to serve as a benchmark for mass-transfer efficiency studies.
\end{abstract}


\maketitle

\section{Introduction}
\label{sec:intro}

A significant fraction of high mass O- and B-type stars are born in binaries and higher-order multiples \citep[e.g.][]{sana_multiplicity_2011, moe_mind_2017, Offner2023}, and interactions between the components are thought to affect a majority of them throughout their lifetime \citep[e.g.][]{sana_binary_2012}. This can shape the physical and chemical evolution of one or both components, through the exchange of mass and angular momentum \citep[e.g.][]{de_mink_rotation_2013}.

The most massive among these stars are likely to end their lives as black holes (BHs) or neutron stars (NSs) following core-collapse supernovae. If the binary survives the explosion, the compact object may remain in a bound system with a stellar companion. If the compact object is a BH and X-ray quiet, it is commonly referred to as a ``dormant Black Hole'' (dBH). Further evolution of the system may lead to an X-ray binary, a second detonation, and/or a binary compact merger \citep[e.g.][]{abbott_observation_2016}. 

This intermediate phase when the compact object is bound to a luminous stellar companion is thought to be comparatively long-lasting, and thus models suggest the presence of O($10^5$) of such systems in the Milky Way Galaxy \citep[e.g.][]{breivik_revealing_2017, langer_properties_2020, nagarajan_realistic_2025, wagg_charting_2026}. Their observational signatures are, however, challenging to detect, since the compact object must be inferred from the motion it induces on the luminous companion, and systems containing two non-compact stars can produce similar observables \citep[e.g.][]{bodensteiner_is_2020, el-badry_unicorns_2022, el-badry_not_2020, Shenar2020b}. These ``imposter'' systems contain two stellar non-compact objects with one of them evolving though a short-lived bloated phase following substantial mass loss. In this phase, the star appears much brighter than expected given its mass and can outshine its companion, leading to the binary being incorrectly identified as a BH-star or NS-star system. 

Despite ongoing efforts, only a few dBHs have been discovered in the Galactic field \citep[e.g.][]{el-badry_sun-like_2023, el-badry_red_2023, gaia_collaboration_discovery_2024} and globular clusters \citep{giesers_et_al_2018,giesers_et_al_2019, whitaker_et_al_2026}. Among massive-star binaries, the only known OB+dBH systems VFTS 243 in the LMC \citep{shenar_x-ray-quiet_2022} and HD 130298 in the Milky Way \citep{mahy_identifying_2022} demonstrate that such systems can be identified through spectroscopic and orbital signatures.

For the Magellanic Clouds, the smaller overall stellar mass implies a lower total number of dBHs, although their lower metallicity leads to weaker stellar winds and hence favours the formation of more massive BH remnants \citep[e.g.][]{mapelli_dynamics_2013}. Additionally, the Magellanic Clouds host a substantial population of young massive clusters, which may provide favourable formation sites for dBHs, e.g. via dynamical interactions. Interestingly, the most massive dBH known in the Milky Way, Gaia-BH3  \citep[33~\msun,][]{gaia_collaboration_discovery_2024}, likely formed via a dynamical interaction in the low-metallicity (and now dissolved) cluster ED-2 \citep{marin_pina_2024}, and thus poses an example of the massive BH population that can result from dynamical low-metallicity environments. Thus, the presence of young massive clusters, and the lower metallicity in the Clouds make them promising targets for dBH searches.

NGC 1850 - BH1 was initially proposed by \cite{saracino_black_2022} as a system containing a main sequence turn-off primary orbiting an unseen secondary. Situated in the LMC cluster NGC 1850, the object's location in the CMD and the cluster's well-constrained age placed initial estimates of the primary mass at around 5 \msun, with simultaneous modelling of the radial velocities (RVs) and lightcurve suggesting a secondary mass around 11 \msun. A non-compact object of this mass would be expected to vastly outshine the proposed 5 \msun\ primary, and due to the cluster's age, no stars of 11 \msun\ could be present in NGC 1850. Thus, the companion was labelled a BH.

Soon thereafter \cite{el-badry_ngc_2022} argued that the system's tight orbit of around 5 days makes the assumption of a normal single-star evolutionary track questionable. As the primary mass inferred by \cite{saracino_black_2022} was based on the system's location in the CMD, a divergence from single-star evolution could also affect the proposed mass of the primary, and thus the inferred companion mass. \cite{el-badry_ngc_2022} suggested constraining the mass via the star's Roche Lobe (RL) geometry and orbit, deriving its mean density, and thus a mass of about 1 \msun. This implies a much lower mass companion, which could feasibly be hidden in the light of the puffed up primary. They further note that such a star could viably be the result of stripping via binary mass transfer, observed in the aforementioned bloated evolutionary stage, after which the primary is expected to contract and heat up again. \citet{stevance_et_al_2022} used BPASS binary population models to predict that a cluster of the mass and age of NGC~1850 should contain a few of such post-interaction systems, though none of the predicted systems matched the turn-off location of NGC1850-BH1.

In a subsequent work, \cite{saracino_updated_2023} updated the optically derived RVs, thus increasing the implied secondary mass. Further, they placed an upper limit of 10\% on the flux contribution of the unseen companion in the MUSE wavelength range, based on spectral disentangling of the MUSE data around the Paschen series. The analysis suggested that even in the case of a substantially stripped primary star, as proposed by \cite{el-badry_ngc_2022}, the implied secondary mass of $\sim$ 5 \msun\ necessitates either a luminous object hidden in an optically thick disk, or a compact companion.

Here, we apply the spectral disentangling technique discussed in \cite{seeburger_autonomous_2024}, based on \cite{simon_disentangling_1994}, to analyse new multi-epoch FLAMES observations of the system in question. These data have a higher spectral resolution than the existing MUSE data and reach further into the blue optical wavelength regime, where the contribution of a luminous companion is expected to be stronger. Using these data, we aim to constrain the nature of NGC1850-BH1 and propose a potential evolutionary history.

\section{Observations and data reduction}
\label{sec:data}

\subsection{FLAMES-ARGUS spectroscopy}

\bh1 was observed with the FLAMES-GIRAFFE spectrograph \citep{2002Msngr.110....1P} installed at UT2 of the ESO Very Large Telescope (VLT) in program 113.26S5.001 (PI: Saracino). The ARGUS integral-field unit (IFU), providing continuous spatial coverage across a field of view of $12\times7$~arcsec with a sampling of 0.52~arcsec/spaxel, was centred on \bh1 to enable us to deblend its spectrum from contributions by nearby stars during the data analysis (cf., Sec.~\ref{sec:analysis} below). Observations were carried out using the LR02 grating, which covers the wavelength range from 3964~\AA\  to 4567~\AA\ at a spectral resolution of R=8\,500. In Table~\ref{table:observations}, we provide an overview of the observations, carried out across 9 individual nights between October and December 2024. Within each night, a single observing block (OB) of two exposures of 1\,350~s exposure time were recorded.

\begingroup 
    \setlength{\tabcolsep}{10pt} 
    \renewcommand{\arraystretch}{1.5} 
    \setlength\extrarowheight{2pt}
    \begin{table}
        \centering
        \begin{tabular}{ c c c c }  
            Date & OB start & OB end & Grade \\
            \hline \hline     
             	4-Oct-2024 & 08:13:14 & 09:13:50 & A \\
	            2-Nov-2024 & 06:42:40 & 07:43:44 & A \\
                24-Nov-2024 & 05:45:19 & 06:44:34 & A \\
	            29-Nov-2024 & 05:36:47 & 06:32:52 & A \\
                1-Dec-2024 & 05:23:52 & 06:23:59 & A \\
                3-Dec-2024 & 03:58:17 & 04:55:11 & A \\
                4-Dec-2024 & 06:23:51 & 07:21:14 & B \\
                6-Dec-2024 & 05:38:30 & 06:37:39 & A \\
                25-Dec-2024 & 03:59:43 & 05:00:17 & A \\   
            \hline \hline  
        \end{tabular}  
        \caption{FLAMES/ARGUS observations of \bh1}
        \label{table:observations}
    \end{table}
\endgroup

We reduced the data using \textsc{P3D} \citep{2010A&A...515A..35S}, a multi-instrument software package for reducing the data of fibre-fed integral-field spectrographs written in \textsc{idl}. \textsc{P3D} performs the different reduction steps (bias subtraction, fibre tracing \& extraction, wavelength calibration, and flat-fielding) on the individual exposures to create a set of raw-stacked spectra (RSS) files. Each RSS file contains a 1D spectrum per spaxel, with all spectra having been resampled to the same wavelength solution. To clean our science exposures from cosmic rays, we utilized the \textsc{PyCosmic} routine \citep{2012A&A...545A.137H}, which is available within \textsc{P3D}.

In Fig.~\ref{fig:argus_fov}, we show a reconstructed 2D image from one reduced exposure. To create the image, we integrated over the central part of the wavelength range and assigned each spaxel to the ($x$, $y$)-location in the focal plane indicated in the ARGUS calibration files available from ESO. The four corners of the ARGUS field of view do not contain fibres. Furthermore, there are three spaxels for which the signal is not recorded by the CCD. Those are visible to the top right of Fig.~\ref{fig:argus_fov}.

\subsection{HST Photometry}
\label{sec:data:hst}

The \textit{Hubble} space telescope (HST) photometry used in this work was published by \citet{2023MNRAS.518.1505K}. It is based on archival WFC3-UVIS observations collected during programs 14069 (PI: Bastian) and 14174 (PI: Goudfrooij) and includes exposures in the filters F275W, F336W, F343N, F438W, F656N, and F814W. The data were analysed with \textsc{Dolphot} \citep{2016ascl.soft08013D} following the approach described in \citet{2019ApJ...887..199G} to create a single photometric catalogue. We matched the final catalogue to \textit{Gaia} data release 3 \citep{2023A&A...674A...1G}, resulting in the recovery of few bright sources that were missing in the HST photometry because they were saturated in all available exposures.

Following the publication of the catalogue in \citet{2023MNRAS.518.1505K}, \citet{2023MNRAS.521.3162S} complemented it with WFC3-IR imaging from program 11913 (PI: MacKenty), resulting in near-infrared photometry in filters F105W, F110W, F125W, F140W, and F160W. The image analysis was analogous to the WFC-UVIS data.

\section{Data analysis}
\label{sec:analysis}

\begin{center}
    \begin{figure}[!t]
        \centering
    	\includegraphics[width=0.5\textwidth]{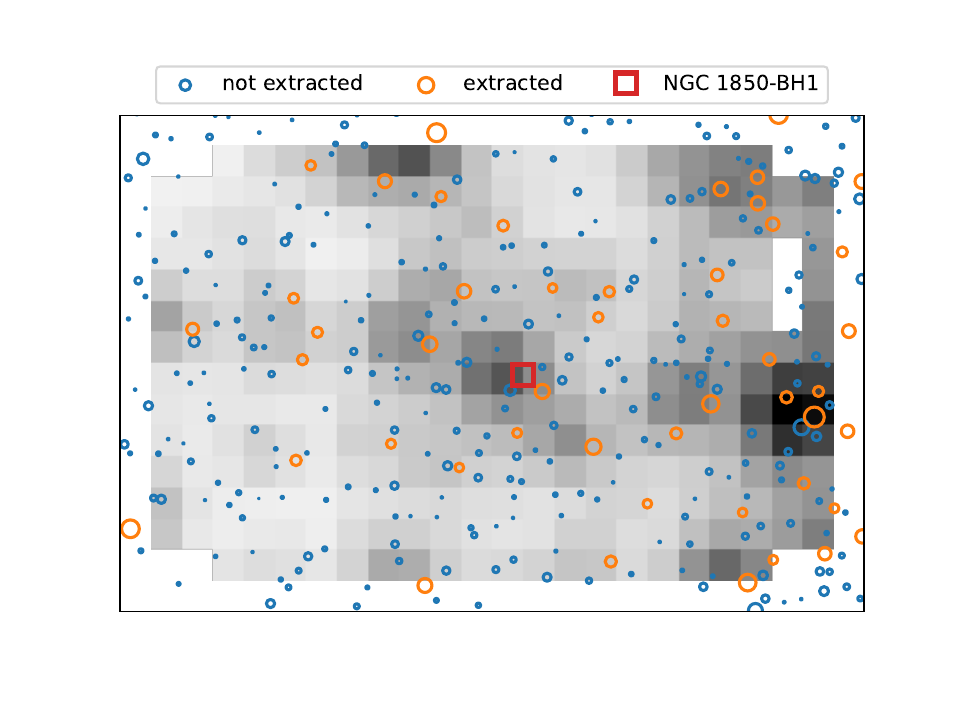}
    	\caption{White light image created from one of the FLAMES/ARGUS exposures. North is \emph{to the right} and east \emph{is up}. Each spaxel covers $0.52\times0.52$~arcsec on the sky. Circles indicate the positions of stars available in the reference HST catalogue, with different colours used for sources for which spectra were extracted (orange) and sources for which no spectra could be extracted (blue). The circle sizes scale with the sources' brightnesses in F438W. The location of \bh1 is highlighted with a red square. Note that the three spaxels without signal visible to the right are attached to fibres where the signal falls outside the CCD footprint.}
        \label{fig:argus_fov}
    \end{figure}
\end{center}

\subsection{Spectrum extraction}

To extract stellar spectra from the individual ARGUS exposures, we used \textsc{PampelMuse} \citep{2013A&A...549A..71K}, a software designed to deblend individual stellar spectra from integral-field spectroscopic observations of crowded stellar fields. \textsc{PampelMuse} requires a reference catalogue of sources within the field of view, which it uses to identify the sources bright and isolated enough to extract their spectra. In our case, we used the HST catalogue described in Sec.~\ref{sec:data:hst} above. To estimate the brightnesses of the stars in the ARGUS data, we adopted the F438W magnitudes available in the catalogue, given the filter's good overlap with the wavelength range of our spectroscopy. In Fig.~\ref{fig:argus_fov}, we indicate the locations of the sources in the HST catalogue relative to the ARGUS field of view as blue circles. Sources for which spectra were extracted are highlighted in orange, while the location of \bh1 is indicated by a red square.

Following the source selection stage, \textsc{PampelMuse} determines the wavelength-dependent point-spread function (PSF) of the ARGUS data and the wavelength-dependent coordinate transformation from the catalogue to the IFU. The former is modelled using a Moffat profile, while the latter uses a six-parameter transformation accounting for any translation, scaling, rotation, and global distortion along either of the spatial axes. With this information in hand, the last step of the analysis is the extraction of the individual spectra.

As discussed in \citet{2022MNRAS.511.2914S}, \bh1 has a neighbour that is only $\sim$0.5~mag fainter, visible as an orange circle to the bottom right in Fig.~\ref{fig:argus_fov}. While we were able to deblend the spectra of the two stars in the ARGUS data, there is a second neighbour, which is about 1.7~mag fainter than \bh1 (visible as a blue circle to the bottom left). This star is too faint and too close to \bh1 to be resolved in the ARGUS data.\footnote{Note that because of the finer spatial sampling and the adaptive-optics enhanced resolution of MUSE, the two sources could be deblended in the data discussed in \citet{2022MNRAS.511.2914S}.} Hence, it is likely that some of the flux from this faint neighbour will contaminate the extracted spectra of \bh1. We will discuss any potential impacts of such contamination on our results in Sec.~\ref{sec:discussion} below.

Following the extraction, we combined the spectra of \bh1 from the two exposures performed within each OB. This resulted in a final sample of nine spectra that were processed further.

\subsection{Spectral disentangling}

To establish the nature of a potential luminous secondary companion (or place upper limits on its maximum flux), spectral disentangling \citep{simon_disentangling_1994, Shenar2020b} is an ideal tool. It uses multi-epoch composite spectra and decomposes them into the individual components, red- and blue-shifted according to their orbital motions. This is particularly useful when the spectra in question are in some way unusual and/or poorly matched by available templates, such as in post-interaction systems. 

For a detailed description of the pipeline employed here, see \cite{seeburger_autonomous_2024}. In short, spectral disentangling assumes that the spectra of the components are time-invariant in their shape (i.e., lines are always the same shape and depth, and are at the same rest wavelength), and that the observed changes in the composite spectrum arise solely from RV shifts of the components. Then, the problem can be framed as asking ``which two spectra can be shifted relative to each other such that they reproduce the observations at all epochs?''. Hence, for a set of trial primary velocities and mass ratios, the secondary velocities can be obtained and the two spectra computed.

Importantly, we can assess the quality of a disentangling result by comparing the observed data to a composite reconstruction from the disentangled components at the relevant shifts for each epoch and computing their \chisq. A small \chisq\, suggests good agreement (within the errors) between observation and reconstruction, and thus a successful disentangling.

To successfully disentangle the spectra, a number of input parameters are needed. These are, at their most basic, the velocities of the primary and secondary component at each epoch. If they are in orbit, then their velocities are related via the mass ratio $q = \frac{M_B}{M_A}$. Thus, the set of input parameters reduces to the primary velocities and the mass ratio.

An initial guess of the primary velocities can be acquired by cross-correlation with a suitable template, or from an existing orbital solution. Here, it does not matter significantly if the template-data match is imperfect, as velocities will be refined in a later step. We use the orbital solution from \cite{saracino_updated_2023} to compute the primary RVs at the dates our spectra were taken.

Holding these fixed, we can now determine an initial guess of the mass ratio. To do so, we trial different values of $q$ and perform spectral disentangling, recording the \chisq\ achieved by the reconstructed spectra for each $q$. Our starting guess is then the $q$ with the lowest associated \chisq.

\begin{figure}
    \centering
    \includegraphics[width=\linewidth]{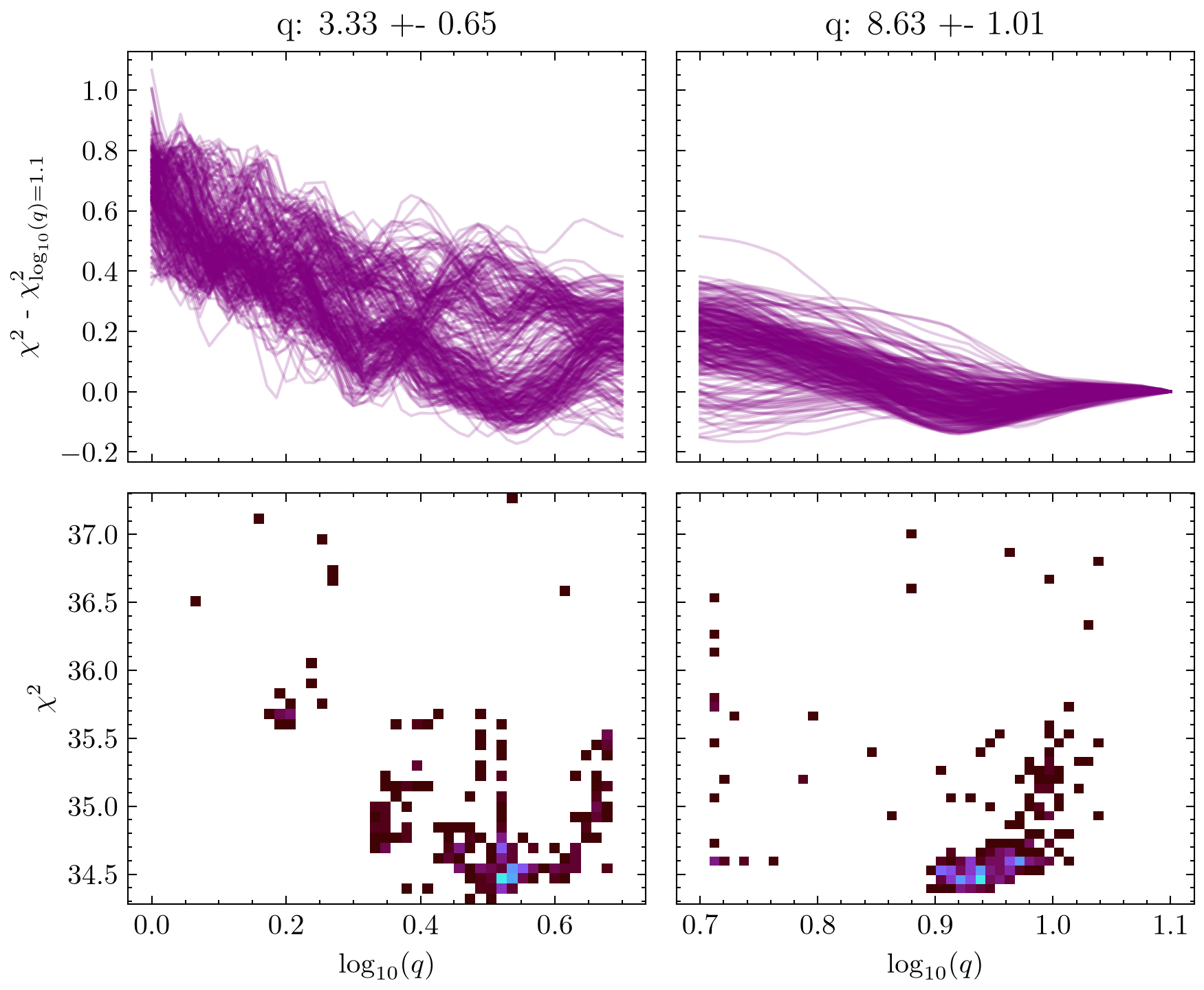}
    \caption{Top panels: \chisq\ values obtained from spectral disentangling as a function of the logarithm of the mass ratio, $\log{q}$, with different lines showing different orbital parameter assumptions. To account for global offsets in the \chisq\ levels of the assumed orbits, we normalise the \chisq\ values to a common value at $\log{q}=1.1$. Bottom panels: 2D histograms of the minimum (absolute) \chisq\ vs. associated $q$ for each realisation. The left panels show $\log{q}$ roughly from 0-0.7, right from 0.7-1.1.}
    \label{fig:q_dist}
\end{figure}

We show this in Fig.~\ref{fig:q_dist}, for a total of 300 orbital realisations. These realisations were generated by sampling from the orbital fit by \citet{saracino_updated_2023}, assuming Gaussian-distributed uncertainties in period, eccentricity, RV semi-amplitude and systemic velocity. Thus, for each realisation we can compute a set of predicted primary RVs. We hold these fixed, and attempt disentangling at a range of values of $q$, recording each solution's \chisq. The top panels show the behaviour of \chisq\ as a function of $\log{q}$, with each line representing one realisation. Note that we normalise the \chisq\ curves to a common value of 1 at $\log{q}=1.1$ to account for the fact that not all orbit realisations fit the data equally well. We split the range into $0 < \log{q} < 0.7$ and $0.7 < \log{q} < 1.1$, as initial explorations revealed two minima in the \chisq\ curves, one within each of the stated ranges. The implications of this \chisq\ behaviour are twofold. First, presence of defined minima suggests that the analysis recovers two luminous companions in the data. Second, the presence of \emph{several} minima indicates that the mass ratio of the two components is not well constrained. To account for the latter, we select the $\log{q}$ with the smallest associated \chisq\ within either sub-range included in Fig.~~\ref{fig:q_dist}. The bottom panels show, for each sub-range, a 2D histogram of these best-fit $\log{q}$ and their associated \chisq. The titles at the top give the mean and standard deviation of the samples, weighted according to their associated \chisq\ value. Subsequently, we demonstrate the analysis using $q = \disq$, but perform the whole pipeline using both values, with a comparative discussion of both values in \S~\ref{sec:discussion}.

After determining initial guesses for RVs and $q$, we next perform \chisq-minimisation, allowing all primary velocities and the mass ratio to vary. This should return the set of parameters that allow the disentangler to most accurately decompose the spectra into their components. This step leads only to small changes in the velocities and mass ratio, of the order of a few percent.

\begin{figure}
    \centering
    \includegraphics[width=\linewidth]{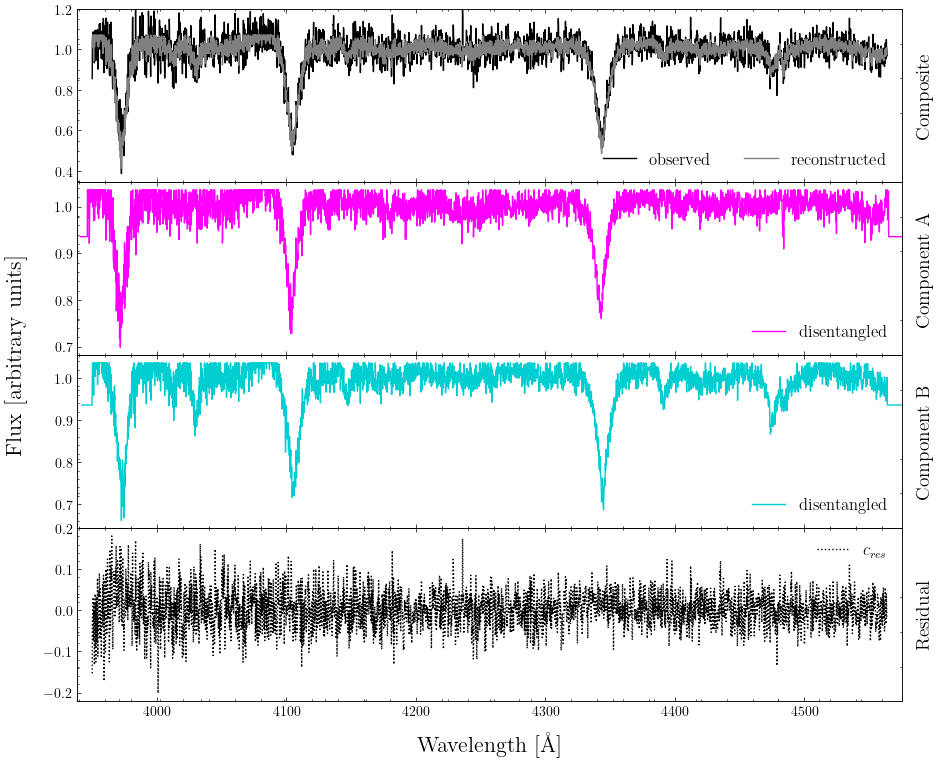}
    \caption{Disentangled FLAMES spectra shown at one epoch (MJD = 60638). The top panel shows the observed spectrum (black), as well as the reconstruction obtained by shifting the disentangled spectra to the relevant velocities and co-adding (grey). The two middle panels show the disentangled spectra of the two luminous components in magenta and cyan. The bottom panel shows the residual, i.e., the difference between observation and reconstruction at the displayed epoch.}
    \label{fig:disentangling}
\end{figure}

The initial results of this process are shown in Fig.~\ref{fig:disentangling}. The top panel shows the composite for one epoch (observation in black, reconstruction in grey), the middle panels the disentangled components, and the bottom panel the residual.
We note the most obvious result, namely the clear presence of two luminous components in the two middle panels. We see the Balmer series present in both stars, identifying them as likely of BAF spectral type. Additionally, we observe MgII at 4481 \AA\ in both the magenta and cyan spectra, as well as Helium-I lines at  4009, 4026, 4121, 4144, 4388 and 4471 \AA\ in the cyan spectrum. Collectively, these place the magenta spectrum as a likely late B- or early A-type star, and the cyan as a potential mid B-type star. Further, the absence of obvious lines or other strange features in the residual indicates that disentangling has likely been successful, and the residual consists primarily of noise. 

\begin{figure}
    \centering
    \includegraphics[width=\linewidth]{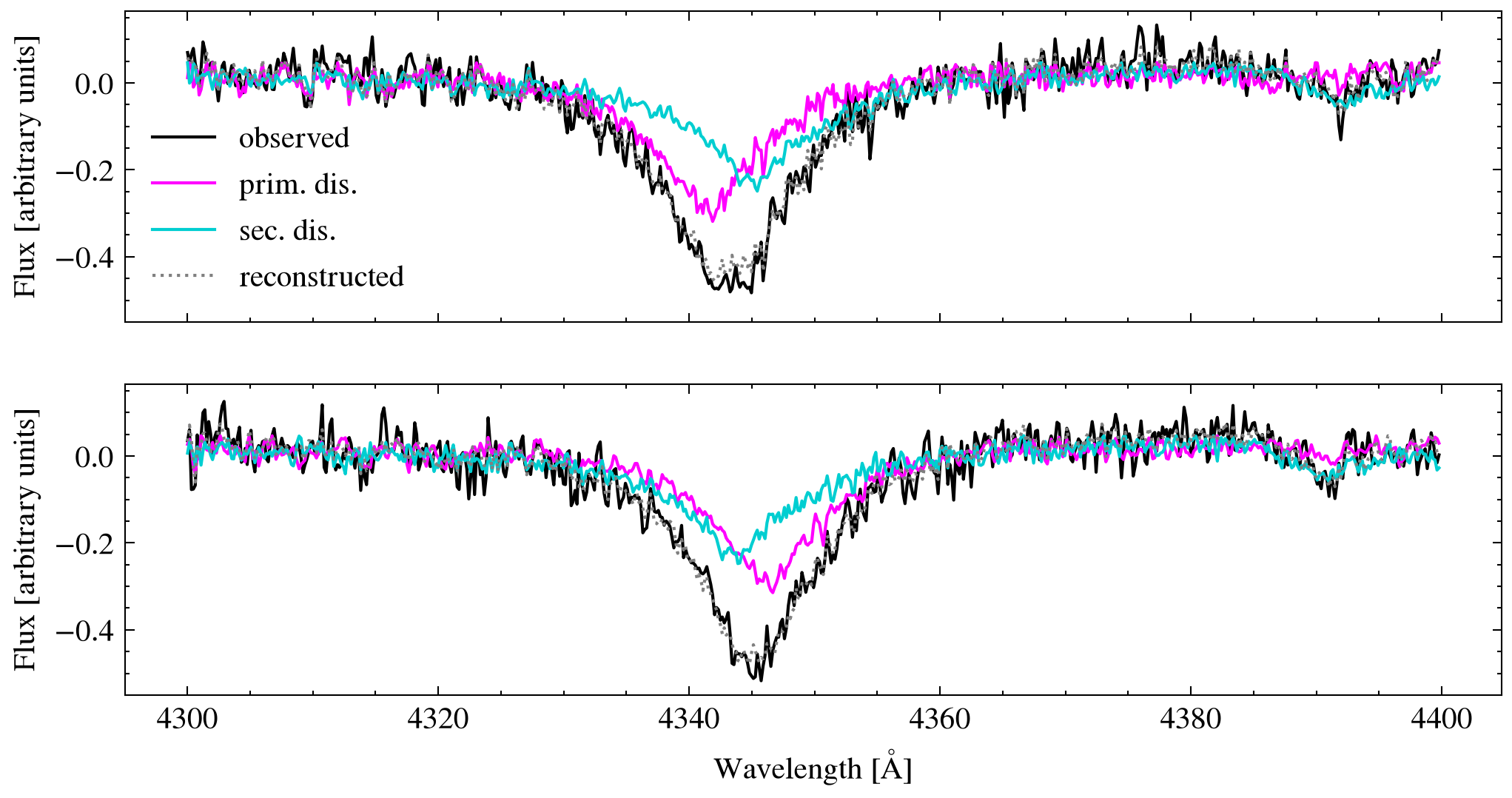}
    \caption{Disentangled FLAMES spectra shown at two epochs around the H$\gamma$ line. The observed spectrum is shown in black, with the primary and secondary disentangled components in magenta and cyan, respectively. Co-adding these disentangled spectra yields the reconstructed epoch spectra, shown with a grey, dotted line.}
    \label{fig:twoepochs}
\end{figure}

We can also inspect the component spectra and associated reconstruction at different epochs, shown in Fig.~\ref{fig:twoepochs}. Here, each panel shows a different epoch, with the observed composite spectrum in black and the disentangled components in magenta (primary) and cyan (secondary). Co-adding them yields the reconstructed spectrum in grey (dotted line). Generally good agreement between the black observation and grey reconstruction increases our confidence in the disentangling result, especially considering the substantial velocity shifts of the components, visible by eye, between the two epochs.

An additional result of the disentangling is a dynamical estimate of the mass ratio $q$, as the secondary velocities at each epoch are calculated from the primary velocities and the value of $q$ which minimises \chisq\, determined during the disentangling process. Here, we find $q$ = \disq, i.e. a secondary that is \disq\, times heavier than the primary. Usually, in binary stars, we use ``primary'' to refer to the more massive, more luminous component of the system - here, the luminosity ratio between the two components is wavelength dependent. We thus elect to refer to the \emph{initially} more massive star as the primary.

\subsection{Template matching}
\label{sec:analysis:templates}

Once the components of the binary have been identified, it is much simpler to match them to available template spectra to derive stellar parameters such as effective temperatures, surface gravities and projected rotation rates. At the same time, we can also determine the flux ratio $f$ between the templates in the relevant wavelength range. This works because the disentangled component spectra have not (yet) been scaled according to a known flux ratio. We thus perform the template fitting by selecting a section of the disentangled component spectra, and computing the total \chisq\, between sets of two templates, scaled by a trial flux ratio. For this step, it is important to note that we expect disentangled spectra to be in the rest frame, allowing consistent comparison with templates.

Here, we use synthetic spectra from \citet{allende_prieto_collection_2018}. The spectra are split into 5 libraries; relevant for this work in particular are libraries 3 and 4. For library 3, \Teff\ ranges from 7000 K to 12000 K in steps of 500 K, while \logg\ ranges from 2 to 5 in steps of 0.5. For library 4, \Teff\ ranges from 10000 K to 20000 K in steps of 1000 K, and \logg\ from 3 to 5 in steps of 0.5. 
Due to the known location of the system in NGC~1850, and to reduce the parameter space, we fix [Fe/H] at -0.5, [$\alpha$/Fe] at 0 and $\log{\xi}$, the logarithm of the microturbulence, at 0.3. We produce our own set of rotation-broadened spectra using \verb|RotBroadInt| \citep{carvalho_simple_2023}, with \vsini\ ranging from 0 km/s to 300 km/s in steps of 10 km/s for each spectrum in the aforementioned libraries.

The synthetic spectra are provided at a resolution of 10000, which we artificially degrade to FLAMES' resolution of $\sim$ 8000 by convolving with a Gaussian. Additionally, we resample the spectra onto the FLAMES wavelength grid, and perform normalisation the same way as on the observed spectra to ensure coherence.

To ascertain uncertainties and increase the robustness of our results, we perform the fitting using Monte Carlo. We do so by creating multiple noise realisations of the templates and repeating the fitting procedure for each realisation. From the set of best fit templates for each realisation, we get the stellar parameters of each component, as well as the flux ratio between them. We select the wavelength ranges and spectral lines that are most sensitive to the parameters we are trying to fit and refine the fit in turn as explained in the following paragraphs. For clarity, we will show the procedure for one noise realisation as an example.

\begin{figure}
    \centering
    \includegraphics[width=\linewidth, trim={0 0 0 0.6cm},clip]{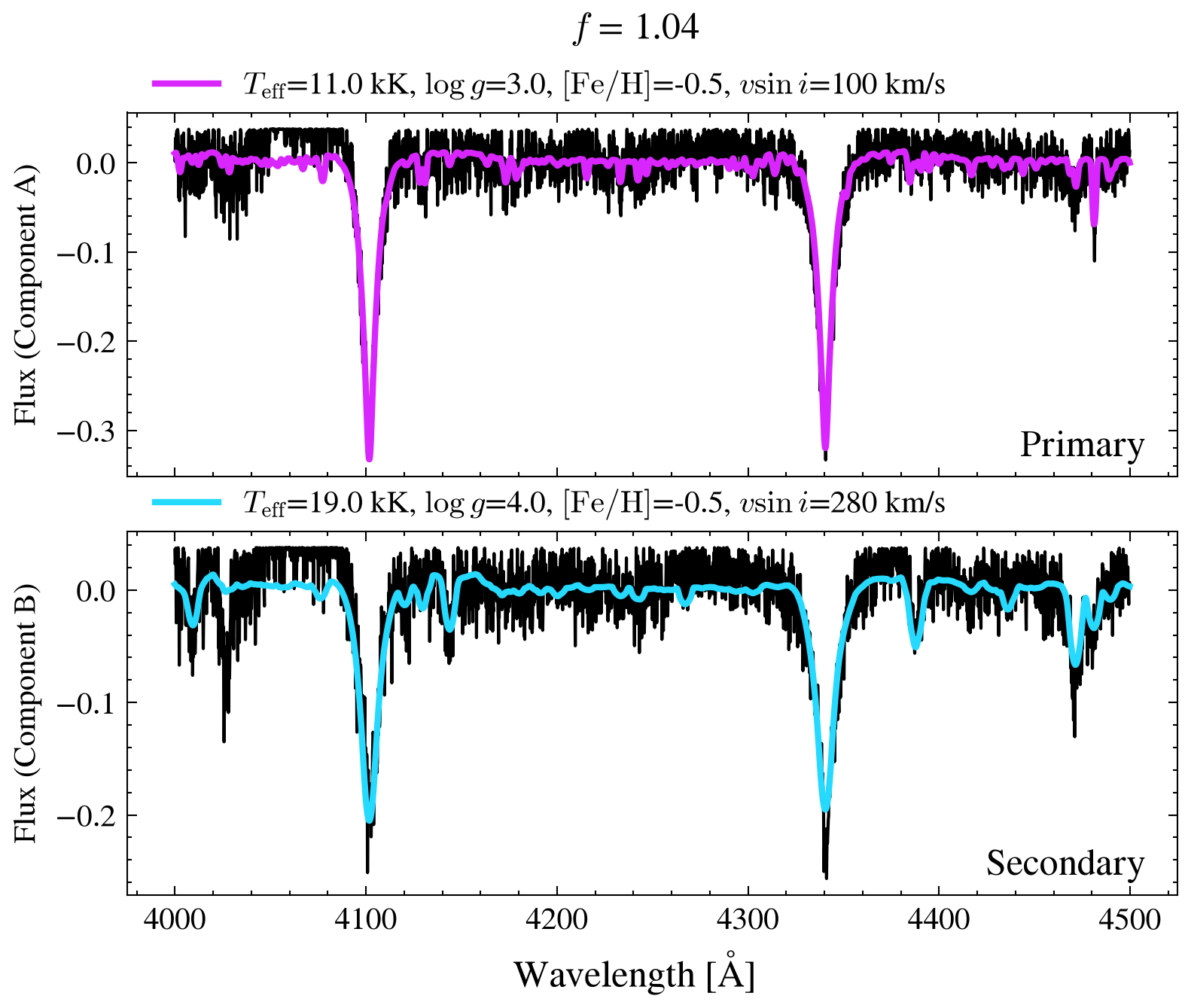}
    \caption{Determining effective temperatures of both components from the best-fit templates. The top panel shows the primary, and the bottom the secondary. Black lines indicate the disentangled spectra, and superimposed magenta and cyan lines the closest matching templates, determined via \chisq minimisation. The values for the various parameters in the legend are for the best-fit template for one realisation, rather than representing the overall best-fit spectrum.}
    \label{fig:teff_flames}
\end{figure}

As the ratios between metal and H/He lines are strong diagnostics of stellar effective temperature, we use almost the entire disentangled spectra to ascertain this. Edge effects can cause spurious solutions here, so we trim off 50 \AA\ on the left and right ends before fitting. Fig.~\ref{fig:teff_flames} shows the disentangled spectra of both components, as well as their closest template match (magenta and cyan, respectively). Visual inspection suggests reasonably good fits in both cases; most lines present in the disentangled spectra are accounted for by the templates, providing constraints on the temperatures \footnote{We note that the Helium-I line at 4026 \AA\ is clearly present in the disentangled secondary spectrum but missing from the template. We assume this is likely due to a poor reproduction of the line in the template.}. The primary (magenta) shows a star with \Teff\ $\simeq$ 11 kK, and the secondary (cyan) with \Teff\ $\simeq$ 19 kK, in agreement with our previous by-eye assessment of the components as a late- and mid-B type stars respectively.

To reliably determine projected rotational velocities, we want to select lines that are minimally affected by shape changes caused by varying surface gravities, disqualifying the Balmer series. Instead, we use MgII at 4481 \AA, present in both stars, to determine \vsini. For the secondary, there is a nearby HeI line at 4471 \AA; we consider this in our fit as well.

\begin{figure}
    \centering
    \includegraphics[width=\linewidth, trim={0 0 0 0.6cm},clip]{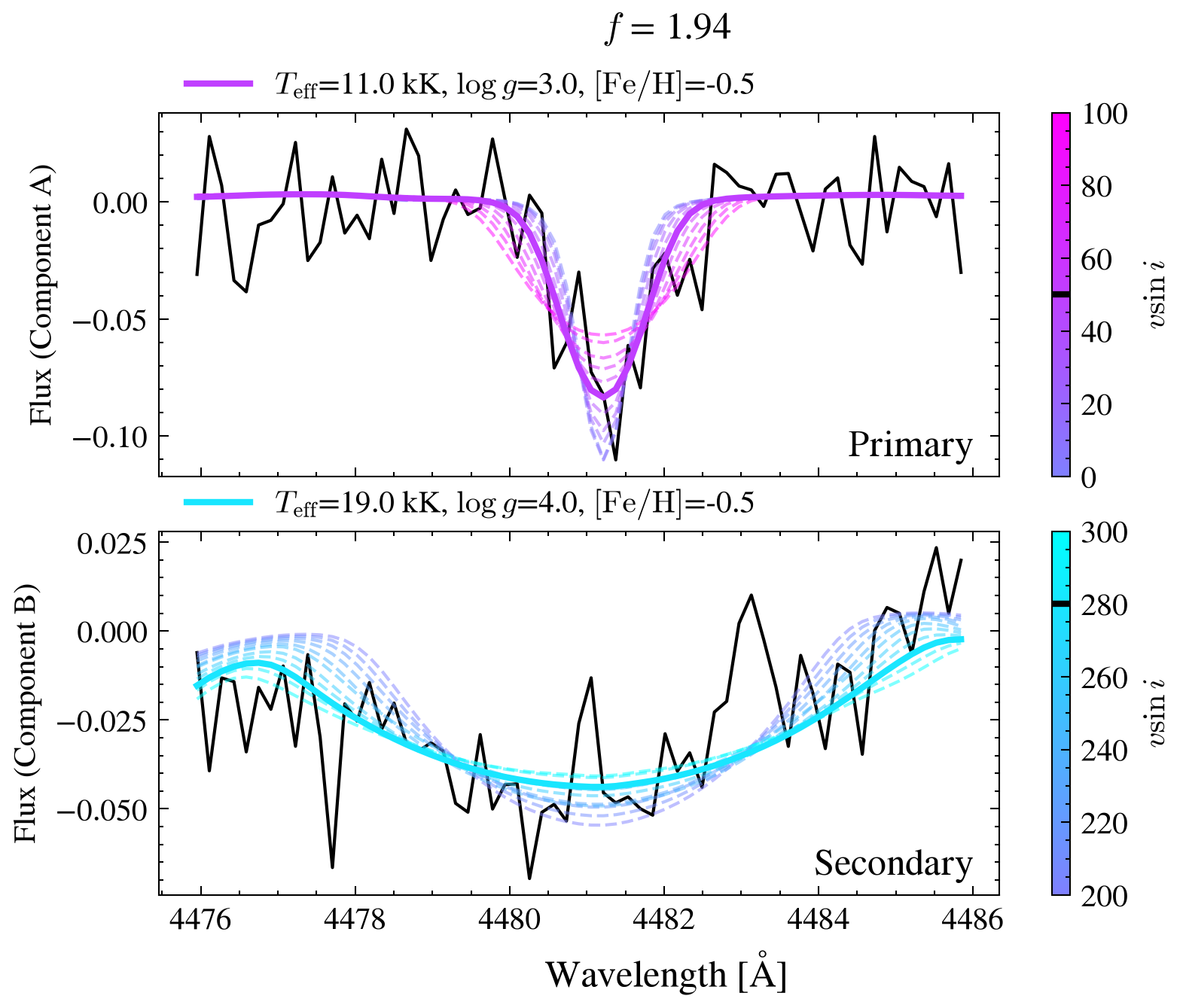}
    \caption{Determining projected rotational velocities for both components from the best-fit templates. Black lines show the disentangled spectra around the MgII line at 4481 \AA, and magenta and cyan dashed lines the considered templates with different \vsini, with their colour indicating the value. The best-fit template is shown using a thicker, solid line, and its associated \vsini\ marked on the colourbar in black. The values for the various parameters in the legend are for the best-fit template for one realisation, rather than representing the overall best-fit spectrum.}
    \label{fig:vrot_flames}
\end{figure}

Next, we hold the effective temperatures of both components fixed at the previously determined values, allowing only \vsini\ to vary and selecting the template with the smallest associated \chisq. Our results for this realisation are shown in Fig.~\ref{fig:vrot_flames}. Here, we find \vsini\ $\simeq$ 60 km/s for the primary, and \vsini\ $\simeq$ 280 km/s for the secondary.

Finally, we wish to determine \Logg\ for both components. For this, the Balmer series is suitable; we perform simultaneous \chisq minimisation for all three Balmer lines present in our disentangled spectra (H$\gamma$, H$\delta$ and H$\epsilon$). The surface gravity primarily affects the shape of the wings of the lines; we determined \vsini\ in a previous step and now hold it constant to avoid degeneracies. 

\begin{figure}
    \centering
    \includegraphics[width=\linewidth, trim={0 0 0 0.6cm},clip]{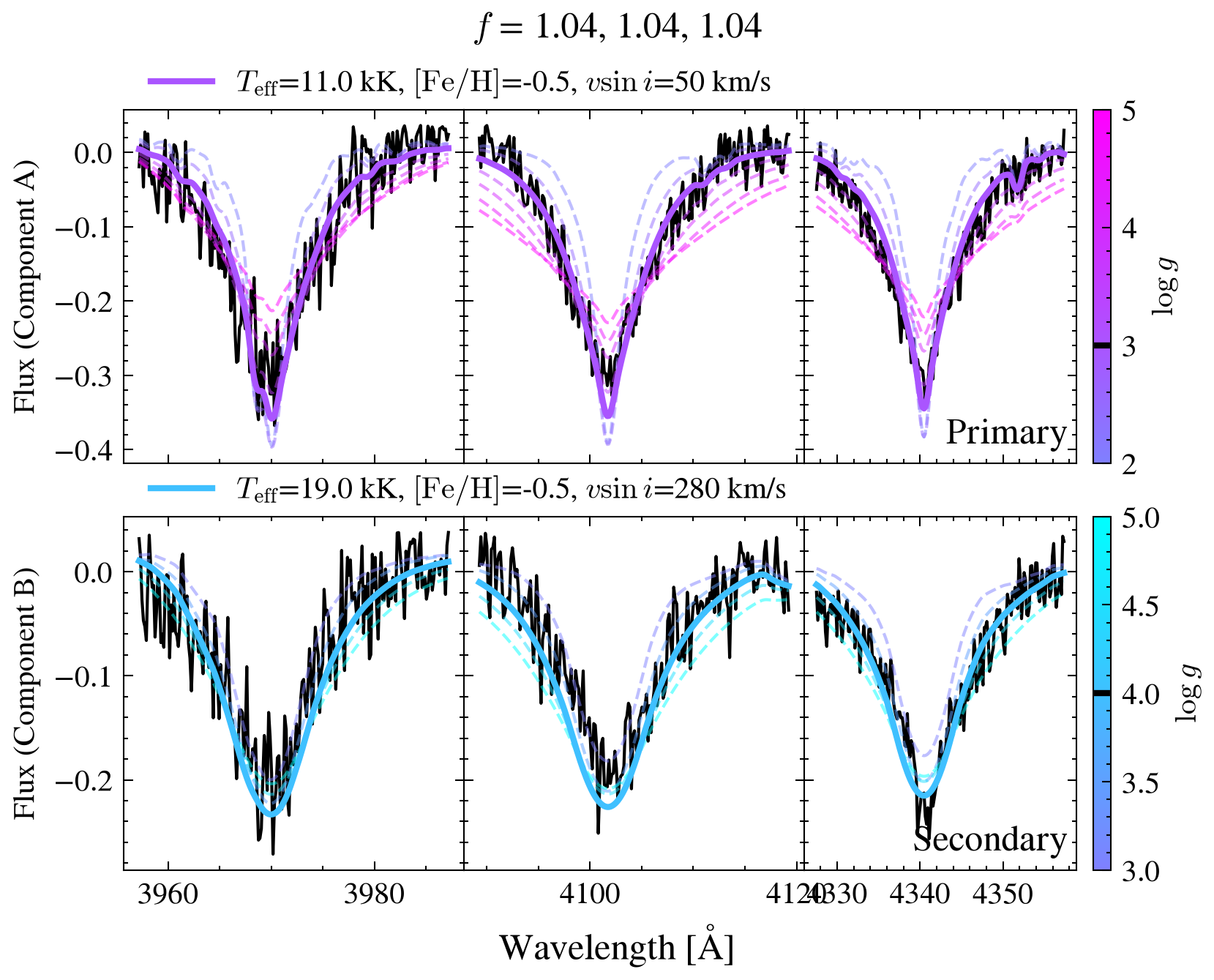}
    \caption{Determining surface gravities for both components from the best-fit templates. Black lines show the disentangled spectra around the Balmer series, and magenta and cyan dashed lines the considered templates with different \Logg, with their colour indicating the value. The best-fit template is shown using a thicker, solid line, and its associated \logg\ marked on the colourbar in black. The values for the various parameters in the legend are for the best-fit template for one realisation, rather than representing the overall best-fit spectrum.}
    \label{fig:logg_flames}
\end{figure}

In Fig.~\ref{fig:logg_flames} we show the result of our \Logg\ fit. For the primary, we determine \logg{A} $\simeq$ 3, and for the secondary, \logg{B} $\simeq$ 4.

As a by-product of the template matching process, we also get a likely flux ratio between the components from the parameter the templates were scaled by. For the lines considered in this work, this was consistently within the errors of a 1:1 flux ratio, i.e. both stars contribute roughly the same amount of flux in the FLAMES wavelength range.

By repeating this process multiple times with different noise realisations, we can take the mean and standard deviation of each assessed parameter to determine likely stellar parameters as well as an error range. These are summarised in the upper part of Table~\ref{tab:params}.

\begin{table}
\centering
\renewcommand{\arraystretch}{1.3}
\caption{The parameters derived for the primary and secondary, using disentangled spectra in conjunction with SED fits.}
\label{tab:params}
\begin{tabular}{llcc}
\hline
Method & Parameter & Primary & Secondary \\ \hline
Disentangling & \Teff [K]      & \TeffA $\pm$ 250 & \TeffB $\pm$ 650 \\
              & \Logg          & \loggA $\pm$ 0.1 & \loggB $\pm$ 0.2 \\
              & \vsini [km/s]  & \vsiniA $\pm$ 14 & \vsiniB $\pm$ 16 \\
 & $q$ & \multicolumn{2}{c}{$3.3 \pm 0.7$} \\ \hline
SED fit       & Radius [$\rsun$] & \radA $\pm$ 0.16 & \radB $\pm$ 0.15 \\
              & \Teff [K]        & 10830 $\pm$ 210 & 18470 $\pm$ 490 \\ \hline
Combined      & Mass [$\msun$]    & \massA $\pm$ 0.4 & \massB $\pm$ 2.6 \\
 & $q$ & \multicolumn{2}{c}{$3.3 \pm 1.8$} \\ 
& \radius{L} [$\rsun$] & \multicolumn{2}{c}{$6.8 \pm 0.5$} \\\hline
\end{tabular}
\end{table}

\subsection{Spectral Energy Distribution fitting}

In addition to spectroscopic data, broad-band photometry can help us further constrain our system parameters and confirm our results. We used the Hubble photometry presented in \S~\ref{sec:data:hst} to fit the components' spectral energy distributions (SEDs). Each star's SED is determined chiefly by its temperature and radius, and attenuated by dust extinction. Assuming we know the distance and E(B-V), we can create a model composite SED consisting of two individual SEDs where we fit effective temperatures and radii of each component until the observations are well matched by our model. We use \verb|pystellibs| \citep{fouesneau_pystellibs_2016} to generate stellar SEDs, the \verb|extinction| package to apply dust attenuation, and \verb|pyphot| \citep{fouesneau_pyphot_2024} to generate mock filter photometry, which we compare to the HST observations.

To perform the fit, we use the Markov chain Monte Carlo sampling package \verb|emcee| \citep{foreman-mackey_emcee_2013}. We set Gaussian priors on the effective temperatures, centred on the values obtained during the template matching of the disentangled spectra described in Sec.~\ref{sec:analysis:templates}. We hold the parallax fixed at its literature value, and place a flat prior on the radii of both components.

By considering our spectroscopic solutions again, we can place further constraints on the SED fit. Specifically, we can compare the spectroscopic flux ratio from disentangling with the flux ratio between the two model SEDs in the wavelength range used for disentangling. This helps break degeneracies in the fitting process and ensures a spectroscopically and photometrically consistent solution. We do this by adding a term to our likelihood function which \verb|emcee| seeks to maximise:

\begin{align}
    \mathcal{L} ={}&
    - 0.5 \cdot
    \left(
        \frac{m_\mathrm{model} - m_\mathrm{HST}}
        {m_\mathrm{error}}
    \right)^2
    \nonumber\\
    &- 0.5 \cdot w \cdot
    \left[
        \log_{10}
        \left(
            \frac{f_\mathrm{don,model} - f_\mathrm{don,data}}
            {\ln(10)}
        \right)
    \right]^2 .
\end{align}.

Here, $m_\mathrm{ model}$ is the mock photometry, $m_\mathrm{ HST}$ the HST observations, and $m_\mathrm{ error}$ their associated errors. $f_\mathrm{ don,model}$ is the ratio of $f_\mathrm{ donor}$ over $f_\mathrm{ total}$ in the simulated model fluxes over the FLAMES wavelength range, and $f_\mathrm{ don,data}$ the flux ratio determined from disentangled spectra. The weight $w$ is chosen to ensure the flux-ratio term does not dominate the likelihood function.

\begin{figure}
    \centering
    \includegraphics[width=\linewidth]{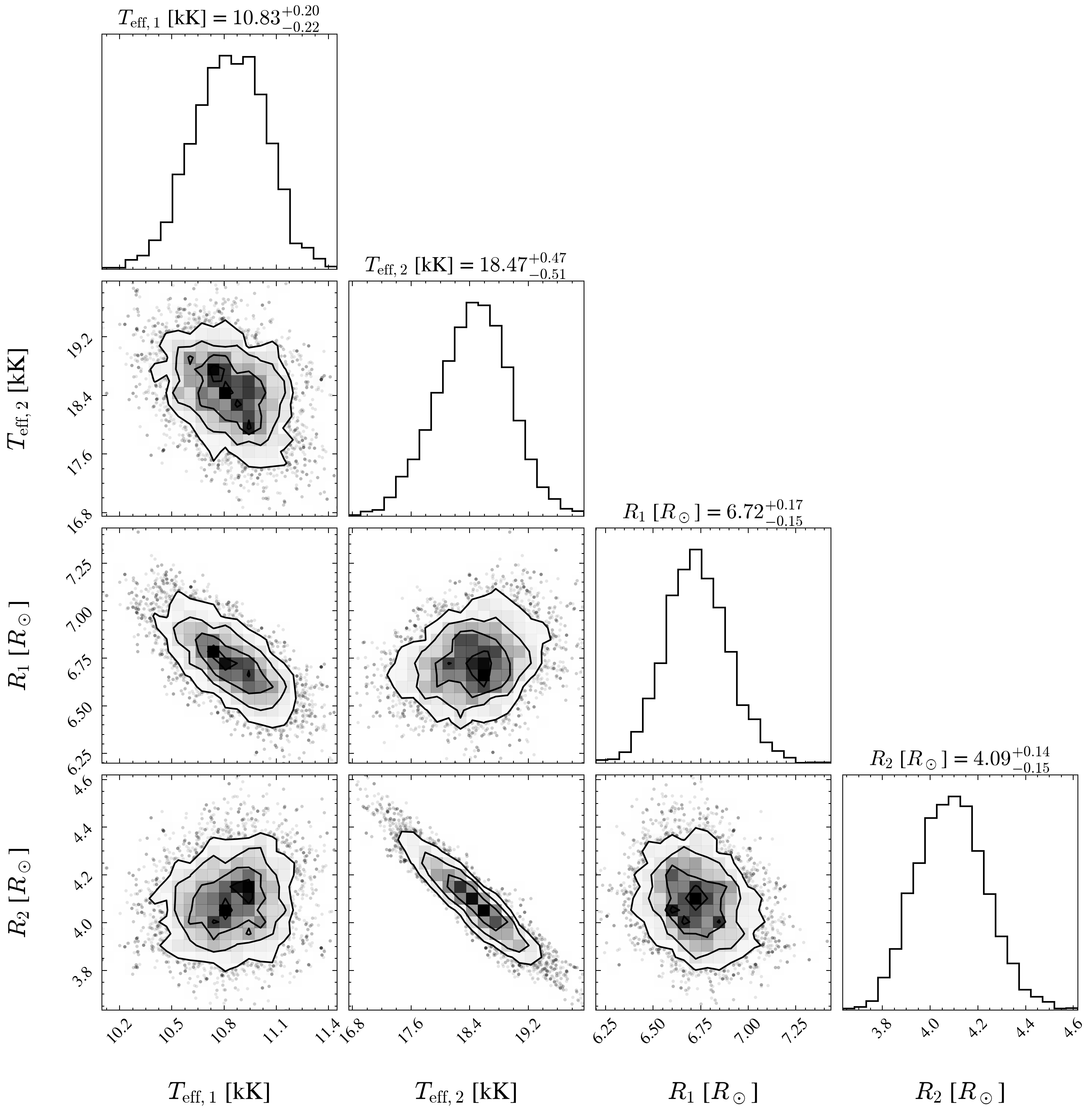}
    \caption{Cornerplot of the parameters fit using the SED of the system.}
    \label{fig:corner}
\end{figure}

Figure~\ref{fig:corner} shows the cornerplot of the parameters (effective temperatures and radii) fit using the SED method. Expectedly, we see a slight anti-correlation between a component's temperature and radius, as both a larger and hotter star lead to a stronger flux, albeit with a different distribution. Overall, the posterior looks well-sampled, strengthening our confidence in our results.

\begin{figure*}
    \centering
    \includegraphics[width=0.7\linewidth]{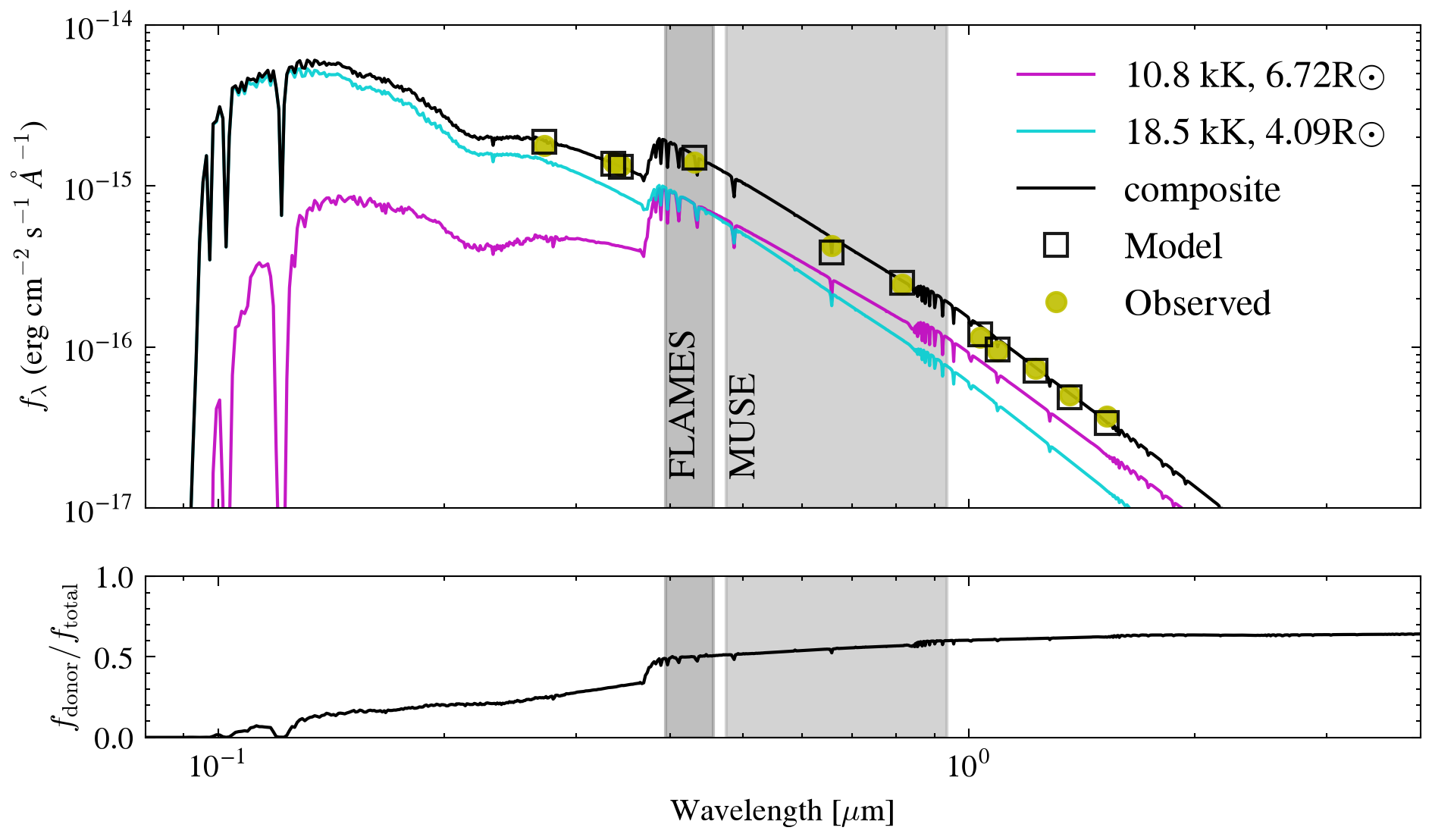}
    \caption{The best-fit two-component SED. \textit{Top panel:} Magenta and cyan lines respectively show the best-fit models for the primary and secondary component, with black the composite. The black boxes show the mock filter photometry for our composite model, while green dots show the HST observations. \textit{Bottom panel:} The black line shows the flux contribution of the donor to the total flux as a function of wavelength, with the red dot marking the flux ratio as determined from spectroscopy at the indicated wavelength. The wavelength ranges observed with FLAMES and MUSE are also shown as grey shaded area in both panels.}
    \label{fig:SED}
\end{figure*}

We show the best-fit SED in Fig.~\ref{fig:SED}, noting good agreement between model (black squares) and observations (green dots). Comparing the model flux ratio to the spectroscopically obtained result also suggests agreement within the errors. The SED fit suggests a primary radius of \radius{A} = \radA\ \rsun, and a secondary radius of \radius{B} = \radB\ \rsun. SED temperatures are consistent within the errors with spectroscopic temperatures - this is not particularly surprising as the spectroscopic temperatures acted as priors on the fit. However, a large discrepancy here would potentially be indicative of an error in the analysis or some other issue, so the congruence is reassuring.

These results can be combined with the previously determined surface gravities of both components to infer stellar masses, providing an additional independent measure of the mass ratio $q$. Here, a primary mass of \mass{A} = \massA\ \msun\ and secondary mass of \mass{B} = \massB\ \msun\ are implied, suggesting a mass ratio of $q$ = 3.3. This is solely computed using effective temperatures and surface gravities from analysing the disentangled spectra, and stellar radii from the SED fit, omitting any dynamical analysis.

With the two component masses and the period as measured by \cite{saracino_updated_2023}, we can compute the primary's Roche lobe using the approximation from \cite{eggleton_aproximations_1983}:

\begin{equation}
    \frac{R_L}{A}
    =
    \frac{0.49q^{-2/3}}
    {0.6q^{-2/3} + \ln\left(1 + q^{-1/3}\right)}.
\end{equation}

Here, $A$ is the orbital separation, $q$ the mass ratio, and \radius{L} the radius of a sphere with the same volume as the Roche lobe. Following this, we get $\radius{L} \sim 6.8 \pm 0.5 \rsun$, implying a primary which is close to filling or fills its Roche lobe. \citet{saracino_black_2022} and \citet{el-badry_ngc_2022} both note that the system is a double-periodic variable \citep[DPV; e.g.][]{mennickent_double-periodic_2003}. OGLE photometry shows both a $P = 5$ d and $P = 156$ d periodicity, with the shorter variability likely due to orbital motion, and the longer related to mass transfer, though its precise origin is not clear. DPVs frequently do not show strong emission lines, despite ongoing mass transfer \citep{mennickent_revealing_2005}.

The determination of these stellar parameters allows us to shed a light onto the evolutionary history of the system, which we discuss in Sec.~\ref{sec:discussion} below.

\section{Comparison with previous results}

This work marks an additional analysis of the target \bh1, and the first such attempt including high-resolution, blue, multi-epoch spectroscopy. We compare and contrast our findings with those of previous studies in Table \ref{tab:comparison}. The properties of the binary components for which this work provides tighter constraints (e.g. masses and flux contributions) are discussed in detail below.

\begin{table*}
\renewcommand{\arraystretch}{1.5}
\centering
\caption{Comparison of the inferred properties of NGC 1850-BH1 in previous
work and in this work.}
\label{tab:comparison}
\begin{tabular}{lllll}
\hline
 & \citet{saracino_black_2022} & \citet{el-badry_ngc_2022} & \citet{saracino_updated_2023} & This work \\
\hline
\mass{A}
  & $\sim 5\,M_\odot$
  & $\lesssim 1.5\,M_\odot$
  & 
  & $1.7 \pm 0.4\,M_\odot$ \\
\mass{A} method
  & CMD/MSTO
  & RL geometry, mean density
  & 
  & \logg{A} \& \radius{A} \\
A State
  & Normal MSTO star
  & Stripped, RL-filling
  & Stripped
  & Stripped, near-RL-filling \\
\mass{B}
  & $\sim 11\,M_\odot$
  & $2.5$--$5\,M_\odot$
  & $\sim 5\,M_\odot$
  & $\sim 5\,M_\odot$\\
\mass{B} method
  & $K_1, P, i$
  & 
  & 
  & \logg{B} \& \radius{B} \\
B nature
  & BH
  & normal MS star
  & unclear
  & accretor \\
$f_{\textrm B}$
  & dark
  & hidden by primary
  & $\lesssim 10\%$ of MUSE flux
  & visible, $\sim$50\% of FLAMES flux \\
$i$
  & $\sim 38^\circ$
  & Higher, diluted ellipsoidal signal
  & $\lesssim 67^\circ$ (no eclipses)
  & N/A \\

\hline
\end{tabular}
\end{table*}

\subsection{Masses}

\citet{saracino_black_2022} derived a primary mass of $\sim$5 \msun\ from the targets location in the CMD. However, this mass estimate did not take into account the possibility of binary interaction, which this work has revealed most likely took place. Mass transfer fundamentally alters the properties of stars, and makes it impossible to use single-star based isochrones to accurately determine stellar parameters. An alternative scenario considering a (partially) stripped donor was originally introduced by \citet{el-badry_ngc_2022}, lowering the primary mass estimate to $\lesssim$ 1.5 \msun\ \citep{el-badry_ngc_2022}. In this work, we find \mass{A} $\sim$1.7 \msun\, consistent with the upper limit of \citet{el-badry_ngc_2022}, with the main uncertainty related to the donor \logg.

The higher primary mass from \citet{saracino_black_2022}, combined with the system's large RV variability and short period lead to a large implied secondary mass of $\sim$11 \msun. As the \citet{el-badry_ngc_2022} interpretation significantly decreases the assumed primary mass, this, in turn, lowers the secondary mass to $2.5 - 5$ \msun. This picture is consistent with this work, where we find a secondary mass of $\sim 5.2 \pm 2.6$ \msun\, with the uncertainty being dominated by the secondary's \logg.

\subsection{Fluxes}

In \citet{saracino_black_2022}, the authors conclude that the secondary is dark, and thus contributes no flux to the system, as no MS star could reasonably fulfill the mass constraints on the secondary while remaining hidden. With the updated masses from \citet{el-badry_ngc_2022}, a luminous secondary ``hiding'' in the light of the puffed-up primary becomes viable. The updated velocities and thus binary mass function of \citet{saracino_updated_2023} imply a secondary that, if it is luminous, contributes at most 10\% of the light in the MUSE wavelength range. They infer that such a secondary must be hidden in an optically thick disc to evade detection in the MUSE disentangling. This work identifies the signature of the luminous secondary, and places its flux contribution at about 50 \% in the FLAMES wavelength range. SED fits paired with spectroscopic temperatures imply a secondary flux contribution between 35-50 \% in the MUSE data, which is in conflict with the 10 \% limit from \citet{saracino_updated_2023}. Possible explanations for this discrepancy include:

\begin{itemize}
    \item An optically thick disc, as potentially implied by IR observations and suggested by \citet{saracino_updated_2023}, which hides (some of) the flux of the secondary in the redder regime. This may also account for the 156 d secondary period. However, the SED fit computed in this work does not necessitate a further component to explain the observed photometry. Additionally, circumstellar material is not expected to selectively absorb at longer optical wavelengths.
    \item As mentioned, \bh1 has a 1.7 mag fainter neighbour, that is too faint and too close to \bh1 to be resolved in the FLAMES data, and thus might contaminate the extracted spectra. While this is possible, and may contribute to the preference for the disentangler for a large mass ratio (assuming this star is single, and thus its RV does not change, at least not on the timescale of the \bh1 orbit), we argue that our recovered secondary spectrum is due to \bh1's companion, rather than this source. First, the neighbour is about 5 times dimmer, while we find our recovered secondary makes up about half the received flux in the FLAMES wavelength range. Second, the neighbour is unlikely to show He-lines, as it is cooler than our recovered secondary. It may, potentially, more strongly affect the MUSE spectra, precisely due to it being cooler and thus more visible in the red. However, as the star could be deblended and a spectrum extracted in the MUSE data, this should not be the case.
    \item Similarly, an unresolved tertiary to the system might contaminate the spectra, as well as induce the additional periodicity.
    \item The accretor as recovered in this work shows only very few, and very broadened lines. In the MUSE wavelength range, we would mainly expect to see H and He lines, where the H lines are also present in the donor star, and thus hard to disentangle. Applying the code used in this work to the MUSE spectra, we find evidence of a secondary with He-absorption at 5016 \AA, but do not manage to recover any other He-lines. Disentangling in the Paschen-line dominated wavelength regime did not reveal a clear signature of a companion, in agreement with the analysis performed by \citet{saracino_updated_2023} in the same regime. One may speculate that the properties of the MUSE spectra (low spectral resolution, non-Gaussian noise properties, only broad Paschen lines) pose a more substantial challenge to spectral disentangling codes than initial tests suggested. 
\end{itemize}

\subsection{Inclination}

\begin{figure}
    \centering
    \includegraphics[width=\linewidth]{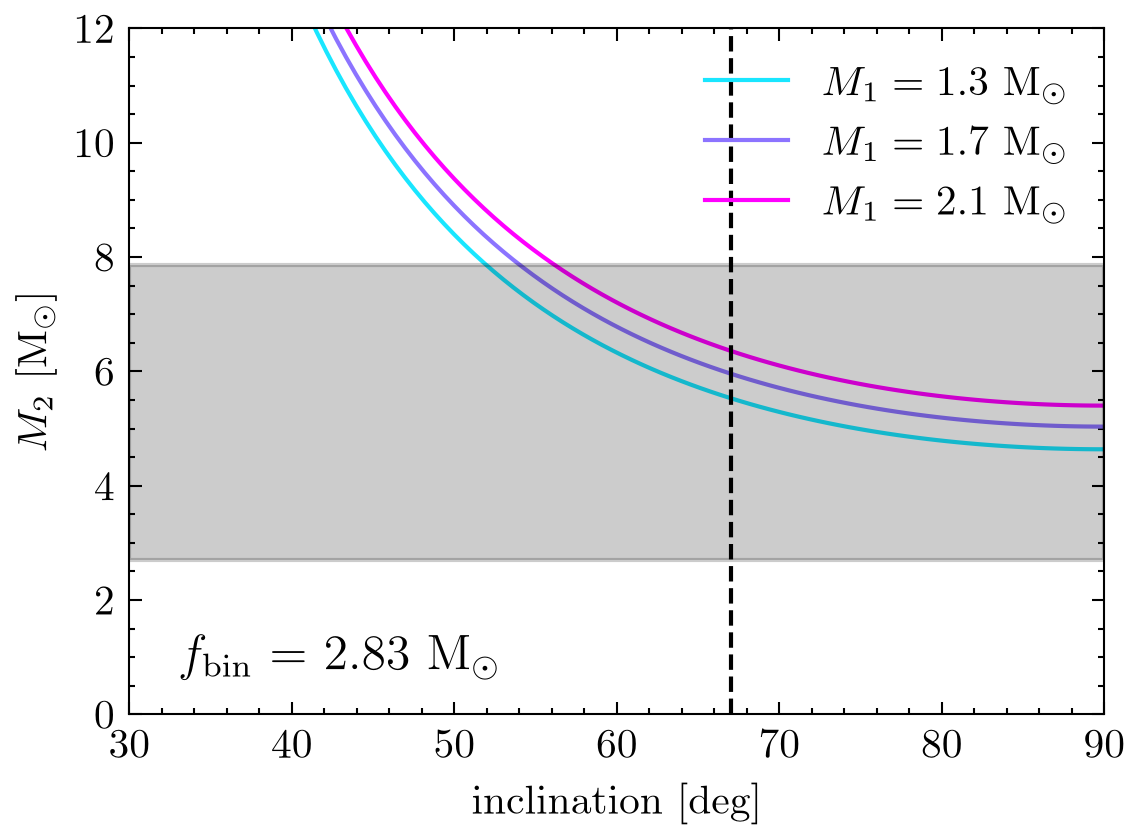}
    \caption{Secondary mass as a function of inclination for a range of primary masses, using the observed binary mass function $f_{\rm bin}$. The different coloured lines show different primary masses (the mean $\pm 1 \sigma$ from our analysis), and the grey shaded region the $1 \sigma$ range of secondary masses found here. We also show, using a vertical, dashed line, the maximum inclination of 67$^\circ$, as found by \citet{saracino_updated_2023}}
    \label{fig:M2_inc}
\end{figure}

\cite{saracino_black_2022}'s initial work constrained the inclination of the to around $38^\circ$, based on lightcurve data. Subsequent work by \cite{el-badry_ngc_2022} called this into question, specifically the assumption of the visible star being a normal 5~\msun\ MSTO star which fills its Roche lobe, and the systems' photometric variability being undiluted by a secondary. Allowing for larger inclination removes the need for a very high mass secondary. The updated analysis in \cite{saracino_updated_2023} contributes two important points to this discussion. First, the lack of visible eclipses constrains the inclination to $i \lesssim 67 ^\circ$. And second, the newly measured velocities imply a larger velocity semi-amplitude of the primary, and thus a larger binary mass function of $f_{\rm bin} = 2.83$. Here, we explore the consequences of these updates.

Figure \ref{fig:M2_inc} follows the idea presented in \cite{el-badry_ngc_2022}, showing the implied secondary mass for a range of primary masses as a function of inclination, using the binary mass function. Here, we use the larger binary mass function from \cite{saracino_updated_2023}, and show the mean $\pm 1 \sigma$ for the primary mass found in this work (coloured lines). The grey shaded region is the $1 \sigma$ range of secondary masses as found here from spectral and SED fits. Considering the constraint from \cite{saracino_updated_2023}, that $i \lesssim 67 ^\circ$, we note that our combination of masses falls essentially just inside the ``allowed'' region. Tighter constraints on the \logg\ for both components using e.g. higher resolution spectroscopy would be helpful in further exploring the here proposed solution and its agreement with constraints found in other work.


\section{Discussion}
\label{sec:discussion}

\paragraph{Evolutionary history}

As spectral disentangling was able to determine two bright components in NGC-1850 BH1, we can confidently rule out a dark secondary in the system. Nevertheless, the current state and evolutionary history of the binary can help shed light on some of the most poorly understood processes in binary evolution, namely binary mass transfer, angular momentum transport, and stellar rotation.

Qualitatively, the primary likely started off as the initially more massive star, expanding as it reached the MSTO. As it did so, it filled its Roche Lobe, and material overflowed onto the secondary. The secondary thus gained both mass and angular momentum, increasing its rotational velocity. After reversal of the mass ratio (the secondary becoming more massive than the primary), further stable mass transfer led to a widening of the orbit, and eventually the primary no longer filled its Roche Lobe, and mass transfer stopped. We infer the cessation of mass transfer from the lack of emission visible in H$\alpha$ as seen in the MUSE spectra in \cite{2022MNRAS.511.2914S}. Ongoing or recently ended mass transfer would be expected to be accompanied by an accretion disk around the secondary, with emission wings around H-$\alpha$, such as in some objects in \cite{seeburger_physical_2026}. 

Previously, it was assumed that the donor star in such a mass-transfer scenario would contract quickly, evolving into a subdwarf. Recently, however, multiple findings of ``bloated'' or ``puffed-up'' stripped stars \citep[e.g.][]{Shenar2020b, bodensteiner_is_2020, el-badry_unicorns_2022} have caused this to be reconsidered. This intermediary stage, before the stripped star settles into a hot, compact subdwarf, is now expected to last much longer, about 10\% of the He-MS lifetime \citep{dutta_evolutionary_2024}.

NGC-1850 BH1 thus belongs to a class of binary systems which masquerade as containing dormant black holes \citep[e.g.][]{rivinius_naked-eye_2020, jayasinghe_giraffe_2022, jayasinghe_unicorn_2021, liu_wide_2019}, but were all subsequently found to be post-interaction binaries with two luminous components \citep[e.g.][]{Shenar2020b, bodensteiner_is_2020, el-badry_unicorns_2022}. In many of these cases, spectral disentangling has provided the definitive proof for a bright companion, as in this work.

\paragraph{The mass ratio}

Interestingly, initial disentangling work showed two plausible minima in the residual for different values of $q$: $\sim \disq$ and $\sim 8.6$. We rule out the larger value by appealing to stellar evolution:
The system's position at the MSTO constrains the initial mass of the primary to around 5 \msun, with an initially less massive companion. For the system to reach the mass ratio of 8.6, assuming initial masses of 5 and 4.5 \msun, the primary must have lost over 4 \msun\ of material which the secondary accreted. This is questionable, both as we expect the core of a 5 \msun\ star to be approximately 1 \msun, and as it would require mass transfer to be nearly fully conservative. 
Additionally, even if one were to assume $q=8.6$ and proceed through the analysis with this value, the resulting values of surface gravities and radii inferred from template and SED fits, and the thus inferred masses, are inconsistent with the assumed value of $q$. This is shown in Appendix~\ref{ap:ap}.

\begin{figure}
    \centering
    \includegraphics[width=\linewidth]{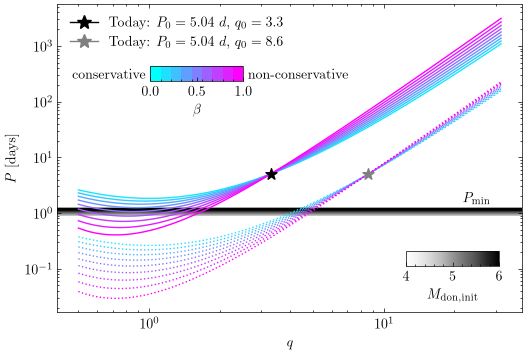}
    \caption{The evolution of the period as a function of mass ratio, calculated backwards from the present properties. We show tracks of different isotropic re-emission parameters $\beta$ (different colours), for two different present-day mass ratios, $\sim 3.4$ and $\sim 8.6$ (black and grey stars and corresponding solid and dashed lines, respectively). Additionally, we show the minimum possible period for the system assuming initial donor masses between 4 and 6 \msun.}
    \label{fig:soberman}
\end{figure}

Finally, appealing to simple equations of angular momentum transfer from \cite{soberman_stability_1997}, we can compute the evolution of the period as a function of mass ratio as we go back in time, as in \cite{bodensteiner_is_2020}. This approach, which is based on the idea that the orbit evolution of a binary depends on the properties of the mass transfer and hence the orbit evolution of the binary can be reconstructed for a given mass-transfer prescription, is summarized in Fig.~\ref{fig:soberman}. Black and grey stars show the present-day period and mass ratio for the two minima found by the disentangler. Corresponding solid and dashed lines illustrate the previous evolution of the period under the assumption of different isotropic re-emission parameters $\beta$. The minimum period is shown with horizontal shading, depending on the initial mass of the donor. Notably, for the larger mass ratio case, all lines, including the one indicating fully conservative mass transfer, dip below this minimum. While it is important to highlight that the \cite{soberman_stability_1997} equations make a number of simplifying assumptions that mean this does not necessarily rule out the $q = 8.6$ case, it is another supporting piece for the $q = \disq$ scenario. In any case, the fact that even in the $q=\disq$ scenario, the lines dip below the minimum allowed period, suggests that future studies of NGC1850-BH1 can be used to stress-test assumptions of binary mass transfer.

\paragraph{Remaining issues and uncertainties}

At this point, it is important to highlight some of the remaining problems and uncertainties with the analysis. 

\begin{itemize} 
    \item The uncertainties associated with the \logg\ values of both components (cf. Table~\ref{tab:params}) are important limitations of our analysis, as the component masses are very sensitive to them. Higher resolution/SNR observations and a more tightly constrained orbital solution might help reduce the errors and better constrain the other parameters in the cascade (masses and mass ratios).
    \item Our templates assume solar-scaled abundances, with a ``normal'' surface hydrogen fraction. However, due to the previous mass transfer, we expect the chemical composition of both components to be altered. Initial exploration (Dutta et al., in prep, using CMFGEN \citep{hillier_treatment_1998}) suggests that a changing surface H-fraction has only small effects on the resultant spectra down to very strong H-stripping, but this may regardless have a knock-on effect on measured \Teff, \logg, \vsini\ and flux ratios.
\end{itemize}



\section{Conclusion}

Spectral disentangling has demonstrated that the system known as NGC 1850 BH1 does in fact not contain a compact companion, but falls into the class of ``imposter'' systems with two bright components. The system presently consists of a luminous, comparatively cool (\teff{A} $\simeq$ \TeffA\ K), puffy (\logg{A} $\simeq$ \loggA) primary, and a more massive, hotter (\teff{B} $\simeq$ \TeffB\ K) and denser (\logg{B} $\simeq$ \loggB) secondary. This atypical configuration is most readily explained by binary interaction. 

This work provides additional analysis, building on that of \citet{saracino_black_2022}, \citet{el-badry_ngc_2022} and \citet{saracino_updated_2023}. While initial work classified the primary as a 5 \msun\ MSTO star and inferred a mass of 11 \msun\ for the (then unseen) companion, subsequent analysis (including this work) has lowered the mass estimate of the primary, revealed the secondary, and placed constraints on its nature, as well as the system's evolutionary history. There are, however, some open questions.

In light of the poorly constrained mass ratio, a full characterisation of the system is still not possible. However, the location of \bh1 inside a cluster of well-known age, metallicity, and distance suggests the system could reveal important insights into our understanding of binary interactions. Therefore, pinning down the mass ratio through observations at higher spectral resolution and/or SNR appears like an important next step. Given the location of the system in a crowded stellar field and the relative faintness of the two components, such observations can currently only be performed from space (e.g., HST-STIS) or with blue-efficient long-slit (e.g., X-Shooter) or integral-field spectrographs at 8m-class ground-based telescopes.

These data might also aid in explaining the discrepancy between the 10\% flux limit on the secondary in the MUSE wavelength range found by \citet{saracino_updated_2023}, and the here implied 30-50 \% flux contribution in the same range. Whether this inconsistency is due to the limited MUSE spectral resolution and similarity of the two spectra in this wavelength regime, dilution from an unrelated star or a tertiary companion, or contamination from a circumbinary disc can only be determined with high-resolution multi-epoch data in the red and/or NIR. Such data may also be useful in exploring the origin of the 156 d secondary period, and provide additional constraints on the parameters of the system and its components.


\section*{Acknowledgments}

Based on observations collected at the European Southern Observatory under ESO programme 113.26S5.001.
RS and SK gratefully acknowledge funding from UKRI through a Future Leaders Fellowship (grant no. MR/Y034147/1). SS acknowledges funding from the European Union under the grant ERC-2022-AdG, ‘StarDance: the non-canonical evolution of stars in clusters’, Grant Agreement 101093572, PI: E. Pancino. TS acknowledges support from the European Research Council (ERC) under the European Union's Horizon 2020 research and innovation programme (grant agreement 101164755/METAL) and the Israel Science Foundation (ISF) under grant number 0603225041. JB is supported by an NWO Veni fellowship (VI.Veni.242.199).

We thank Ylva Götberg, Hugues Sana, Matti Dorsch and Ulrich Heber for helpful discussion of the target, as well as support with the original FLAMES proposal.

\bibliographystyle{apsrev4-1}

\bibliography{oja_template}

\begin{appendix}

\section{The $\lowercase{q} = 8.6$ case}
\label{ap:ap}

\begin{figure*}
    \centering
    \includegraphics[width=\linewidth]{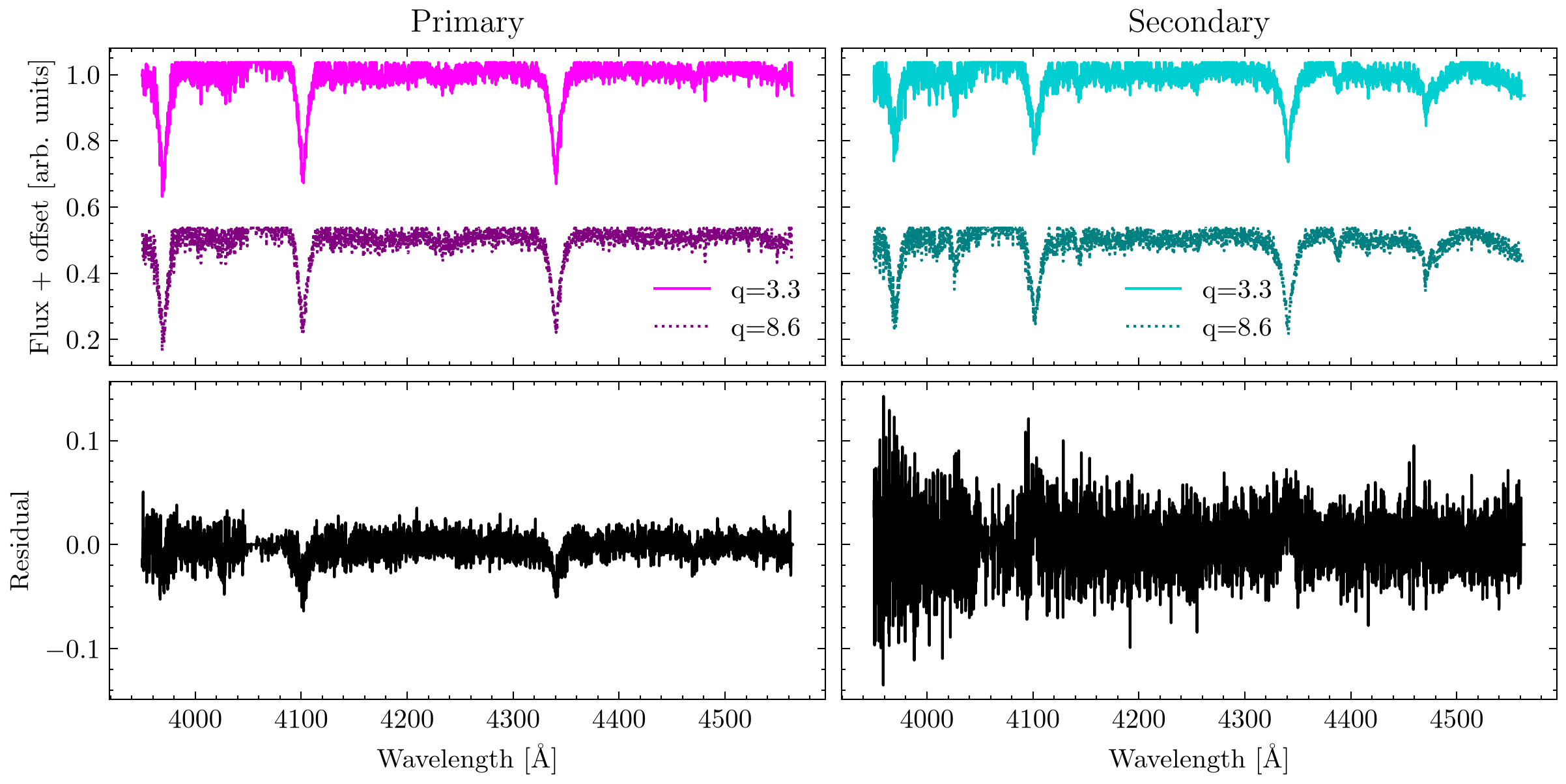}
    \caption{Comparing the disentangling results using two different mass ratios of $q = 8.6$ and $q = \disq$, keeping primary and centre of mass velocities fixed. The disentangled primary and secondary are shown in the top two panels in magenta and cyan, respectively, with the different mass ratios shown, offset, with different shades and lines (lighter, solid for $q = \disq$ and darker, dotted for $q = 8.6$). The bottom two panels show the respective differences between the two solutions.}
    \label{fig:compare}
\end{figure*}

In Fig.~\ref{fig:compare}, we show the different results from disentangling when using the same primary and centre of mass velocities, but different $q$. Qualitatively, the disentangled spectra using $q = 8.6$ are similar to those achieved using $q = \disq$ (top two panels, darker, dotted and ligher, solid lines respectively). Differences become more apparent when looking at the residuals, shown in the bottom panels. Particularly around the Balmer series, the line shape changes slightly, leading to deeper lines in the primary for the $q = \disq$ case over the $q = 8.6$ case. The secondary lines are affected in the opposite way, with the Balmer series deeper for larger mass ratios. This likely happens as a larger mass ratio leads to less computed secondary motion, and thus more of the line depth gets ascribed to the secondary during disentangling. These differences are subdominant to noise when fitting templates, leading to similar (within the errors) stellar parameters, shown in Table \ref{tab:paramsalt}. We note that the parameters generally fall within the errors of those acquired in the $q = \disq$ analysis. The biggest change occurs in the masses and thus the combined-method (spectral \logg\ and SED radii) mass ratio. Even here, however, due to the large errors owing to the comparatively large uncertainty on \logg, the results between the two analyses are consistent with each other. However, we note that the value of $q=3.5\pm2.4$ derived from a combination of SED-based radii and spectroscopically determined \Logg\ is inconsistent with the value of 8.6 obtained from disentangling.

\begin{table}
\centering
\renewcommand{\arraystretch}{1.3}
\caption{The parameters derived for the primary and secondary when assuming an initial mass ratio of $q=8.6$, using the disentangled spectra in conjunction with the SED fits.}
\label{tab:paramsalt}
\begin{tabular}{llcc}
\hline
Method & Parameter & Primary & Secondary \\ \hline
Disentangling & \Teff [K]      & 10500 $\pm$ 280 & 18860 $\pm$ 400 \\
              & \Logg          & 3.0 $\pm$ 0.2 & 3.9 $\pm$ 0.2 \\
              & \vsini [km/s]  & 56 $\pm$ 12 & 278 $\pm$ 16 \\
 & $q$ & \multicolumn{2}{c}{$8.6 \pm 1.0$} \\ \hline
 SED fit       & Radius [$\rsun$] & 6.8 $\pm$ 0.18 & 4.1 $\pm$ 0.11 \\
              & \Teff [K]        & 10720 $\pm$ 200 & 18550 $\pm$ 340 \\ \hline
Combined      & Mass [$\msun$]    & 1.9 $\pm$ 1.0 & 5.4 $\pm$ 2.6 \\
 & $q$ & \multicolumn{2}{c}{$3.5 \pm 2.4$} \\ 
 & \radius{L} [$\rsun$] & \multicolumn{2}{c}{$6.8 \pm 0.5$} \\\hline
\end{tabular}
\end{table}

\end{appendix}

\end{document}